\documentclass[aps,twocolumn,showpacs,nofootinbib]{revtex4-1}

\usepackage{amsmath,amssymb,bm}
\usepackage[dvips]{graphicx}
\usepackage{hyperref}
\usepackage{yfonts}
\usepackage{color}
\newcommand{\beq}{\begin{eqnarray}}
\newcommand{\eeq}{\end{eqnarray}}

\newcommand{\U}{\text{U}}
\renewcommand\d{\partial}
\begin{document}

\title{Generalized Bloch theorem and chiral transport phenomena}

\author{Naoki Yamamoto}
\affiliation{Department of Physics, Keio University,
Yokohama 223-8522, Japan}

\begin{abstract}
Bloch theorem states the impossibility of persistent electric currents in the ground 
state of nonrelativistic fermion systems. We extend this theorem to generic systems 
based on the gauged particle number symmetry and study its consequences on 
the example of chiral transport phenomena. We show that the chiral magnetic effect 
can be understood as a generalization of the Bloch theorem to a nonequilibrium 
steady state, similarly to the integer quantum Hall effect. On the other hand, 
persistent axial currents are not prohibited by the Bloch theorem and they can be 
regarded as Pauli paramagnetism of relativistic matter. An application of the 
generalized Bloch theorem to quantum time crystals is also discussed.
\end{abstract}
\maketitle

\section{Introduction}
During the 1930s Felix Bloch demonstrated the impossibility of persistent electric 
currents in the ground state of interacting nonrelativistic systems \cite{Bohm}.
This Bloch theorem invalidated the idea proposed by Landau and others that 
superconductivity is characterized by persistent ground-state currents \cite{history}; 
see also Ref.~\cite{Ohashi} for its extension to nonrelativistic systems at finite temperature.

Recently, the idea of spontaneous currents has revived in a completely different 
context: the chiral magnetic effect (CME) 
\cite{Vilenkin:1980fu, Nielsen:1983rb, Alekseev:1998ds, Fukushima:2008xe} 
and chiral vortical effect (CVE) \cite{Vilenkin:1979ui, Son:2009tf, Landsteiner:2011cp}. 
As originally argued by Vilenkin \cite{Vilenkin:1979ui,Vilenkin:1980fu},
the CME and CVE are considered the ``ground-state (or equilibrium) currents'' 
in relativistic systems with chirality imbalance in a magnetic field or in a rotation
[see Eq.~(\ref{CME}) below]. Remarkably, they are manifestations of the topological 
nature of chiral fermions, and have a close connection with the topological and 
quantum phenomenon known as the axial anomaly in field theory \cite{Adler, BellJackiw} 
and with the Berry curvature \cite{Volovik, Son:2012wh,Stephanov:2012ki,Son:2012zy,
Chen:2012ca,Manuel:2013zaa,Chen:2014cla,Manuel:2014dza}. These chiral transport 
phenomena are expected to appear in a wide area of physics from condensed matter 
physics \cite{Nielsen:1983rb, Alekseev:1998ds} and nuclear physics 
\cite{Son:2004tq, Fukushima:2008xe} to cosmology \cite{Joyce:1997uy,Boyarsky:2011uy}
and astrophysics \cite{Charbonneau:2009ax, Ohnishi:2014uea, Kaminski:2014jda}, 
and were studied in the framework of gauge-gravity duality \cite{Erdmenger:2008rm, Banerjee:2008th}.

One can ask whether the Bloch theorem can be generalized to apply to the CME and 
CVE and whether they are really possible in the ground state or in equilibrium.%
\footnote{When the magnetic field is \emph{dynamical}, the system with the CME is 
unstable due to the chiral plasma instability \cite{Akamatsu:2013pjd}
(see also Refs.~\cite{Redlich:1984md, Joyce:1997uy, Boyarsky:2011uy}), 
and is not in apparent contradiction with the Bloch theorem. In this paper, 
we shall concentrate the case with the \emph{external} magnetic field.}
This question is also important for possible technological applications of the CME and 
CVE; if electric currents could flow even in equilibrium, one could make best use of 
them without energy loss, in contrast to the Ohm's current that dissipates energy via 
Joule heat; see also Ref.~\cite{Avdoshkin:2014gpa} for a similar question from a different 
perspective and Refs.~\cite{Vazifeh,Burkov,Basar:2013iaa,Landsteiner:2013sja} for related 
issues in the context of Weyl semimetals \cite{Vishwanath,BurkovBalents,Xu-chern}.

The purpose of this paper is to resolve this question as well as to discuss 
other possible applications of the Bloch theorem. To this end, we first extend the 
Bloch theorem to generic systems, including relativistic systems, 
based on the consequence of the gauged $\U(1)$ particle number symmetry.
This indicates that total chiral magnetic currents should vanish in the ground 
state of any system. Moreover, we explicitly show that the CME can be understood 
as a generalization of the Bloch theorem to a nonequilibrium steady state, similarly 
to the integer quantum Hall effect (IQHE) \cite{Laughlin, Thouless}.

We emphasize that the essence of the argument for the generalized Bloch theorem 
is the \U(1) (vector) gauge symmetry. More specifically, our argument is based on the 
fact that the coordinate-dependent space translation (or spatial rotation) of a {\it state} 
can be regarded as the vector gauge transformation of {\it fields} in a given theory.
By using the ambiguity of this gauge symmetry, it can be shown that any current 
flowing state is not the ground state in the thermodynamic limit. 
This is independent of the details of systems and is applicable to any particle number 
currents, not limited to the CME. On the other hand, as there is no such \U(1) axial 
gauge symmetry, spontaneous axial currents in the ground state are {\it not} forbidden 
by the Bloch-type no-go theorem. We indeed show that the spontaneous axial current, 
known as the chiral separation effect (CSE) \cite{Son:2004tq, Metlitski:2005pr}, 
can be understood as Pauli paramagnetism of relativistic matter.

This paper is organized as follows. 
In Sec.~\ref{sec:NR}, we review the original argument of the Bloch theorem and 
its extension by Bohm to circulating currents in nonrelativistic systems.
In Sec.~\ref{sec:extension}, we extend the Bloch theorem to generic systems.
In Sec.~\ref{sec:application}, we discuss its applications to chiral transport phenomena. 
We also comment on the application to the question of quantum time crystals 
proposed by Wilczek \cite{Wilczek:2012jt}. In Sec.~\ref{sec:physics}, we provide 
a physical derivation of the CME and CSE as a nonequilibrium steady current 
and a spin polarization, respectively. 
Section~\ref{sec:conclusion} is devoted to our conclusions.

Throughout the paper, we set $\hbar = c = e=1$ for simplicity unless 
otherwise stated. We will concentrate on systems at zero temperature.

\section{Bloch theorem for nonrelativistic Hamiltonian}
\label{sec:NR}

\subsection{No-go theorem for total ground-state currents}
\label{sec:bloch}
Let us briefly review the original argument of the Bloch theorem for a 
nonrelativistic electron system \cite{Bohm}. The Hamiltonian is given by 
\begin{gather}
\label{H}
H_{\rm NR} = \int d^{3} {\bm x}\, \psi^{\dag}({\bm x})
\left( -\frac{{\bm \nabla}^2}{2m} - \mu \right) \psi({\bm x})
\nonumber  \\
+  \int d^{3} {\bm x}d^{3} {\bm x'}\,
\psi^{\dag}({\bm x}) \psi^{\dag}({\bm x'}) V({\bm x} - {\bm x'})
\psi({\bm x}) \psi({\bm x'}),
\end{gather}
where $\mu$ is the chemical potential and $V({\bm x} - {\bm x'})$ is the isotropic
and homogeneous electron-electron interaction. For simplicity of notation, we here 
omit the spin degrees of freedom, but it is straightforward to generalize the 
argument to electrons with spin \cite{Ohashi}. For later purpose, we also 
introduce the Hamiltonian density ${\cal H}_{\rm NR}$, which is related to 
$H_{\rm NR}$ by 
\beq
H_{\rm NR} = \int d^3 {\bm x} \, {\cal H}_{\rm NR}(\bm x).
\eeq

Let us first assume that the ground state $|\Omega \rangle$ that carries 
a nonzero electric current, $\langle {\bm J}_{\rm NR} \rangle \neq 0$, exists. 
Here and below, the expectation value of an operator ${\cal O}$ with respect to 
the ground state $|\Omega \rangle$ is denoted as $\langle {\cal O} \rangle$.
The total current is defined by 
\begin{gather}
{\bm J}_{\rm NR} = \int d^{3} {\bm x}\, {\bm j}_{\rm NR}(\bm x), \\
{\bm j}_{\rm NR}(\bm x) = \frac{1}{2im}(\psi^{\dag} {\bm \nabla} \psi - \psi {\bm \nabla} \psi^{\dag}).
\end{gather}
By definition, the ground state $|\Omega \rangle$ minimizes the total energy, 
$\langle H_{\rm NR} \rangle \equiv  \langle \Omega | H_{\rm NR}|\Omega \rangle = E_{\rm NR}^{\rm min}$.

We now consider the trial state, 
\beq
\label{GS}
|\Omega' \rangle=e^{i \delta {\bm p} \cdot {\bm x}}|\Omega \rangle,
\eeq
with the momentum $\delta {\bm p}$ being arbitrary at this moment.
Taking the expectation value of $H_{\rm NR}$ for the trial state $|\Omega' \rangle$, 
one finds that the potential energy does not change while the kinetic energy does.
The total energy is given by
\begin{gather}
\label{E}
E'_{\rm NR} = E_{\rm NR}^{\rm min} + {\delta} {\bm p} \cdot \langle {\bm J}_{\rm NR} \rangle + 
\frac{({\delta} {\bm p})^2}{2m}\langle N \rangle,  \\
N = \int d^{3} {\bm x}\, n({\bm x}), \qquad n({\bm x}) = \psi^{\dag} \psi,
\end{gather}
where $E'_{\rm NR} \equiv \langle \Omega' | H_{\rm NR}|\Omega' \rangle$.

As we assumed that $\langle {\bm J}_{\rm NR} \rangle \neq 0$, if we choose
the magnitude of $\delta {\bm p}$ infinitesimally small so that the third term 
on the right-hand side of Eq.~(\ref{E}) is negligible, and if we choose its direction 
opposite to $\langle {\bm J}_{\rm NR} \rangle $, we have $E'_{\rm NR} < E_{\rm NR}^{\rm min}$.
However, this contradicts the original assumption that the ground state has the 
lowest energy. Therefore, one concludes that $\langle \bm J_{\rm NR} \rangle \neq 0$ 
is forbidden in the ground state. This completes the proof of the Bloch theorem.

\subsection{No-go theorem for circulating currents}
\label{sec:bohm}
The above result itself does not forbid the presence of a ground-state 
circulating current, since its integral over space is vanishing. 
As shown by Bohm for nonrelativistic systems \cite{Bohm}, however, the 
Bloch theorem can also be extended to such circulating currents in the 
thermodynamic limit. For completeness of the paper, we recapitulate Bohm's 
result in this subsection.

We consider a ring with the width $\Delta r$ at radius $r$ ($\Delta r \ll r$) 
in polar coordinates $(r, \phi)$, and we shall take the thermodynamic limit 
($r \rightarrow \infty$ with $\Delta r$ fixed) in the end.
We define the circulating current and the energy  as 
\begin{align}
\label{j_circle}
{\cal J}_{\rm NR} &\equiv \oint_C {\bm j}_{\rm NR} \cdot d {\bm l} = 2\pi r j_{\rm NR}, \\
\label{E_circle}
{\cal E}_{\rm NR} &\equiv \oint_C \langle {\cal H}_{\rm NR} \rangle \, dl 
= 2\pi r \langle {\cal H}_{\rm NR} \rangle,
\end{align}
where the line integral is taken along the circle with the radius $r$, and 
\beq
\label{j_local}
j_{\rm NR}({\bm x}) =-\frac{i}{mr} \psi^{\dag}({\bm x}) \frac{\d}{\d \phi} \psi({\bm x})
\eeq
is the current density operator in polar coordinates. 
The total current, energy, and number of fermions on the ring are given by 
$J_{\rm NR} = {\cal J}_{\rm NR} \Delta r$, $E_{\rm NR} = {\cal E}_{\rm NR} \Delta r$, 
and $N = 2\pi r \Delta r \langle n \rangle$, respectively.
We denote the ground state by $|\Omega \rangle$, which has the lowest 
energy, $E_{\rm NR}=E_{\rm NR}^{\rm min}$, or 
${\cal E}_{\rm NR}={\cal E}_{\rm NR}^{\rm min}$ when divided by $\Delta r$. 

Let us consider the total energy of the trial state,
\beq
\label{GS2}
|\Omega' \rangle=e^{i k \phi}|\Omega \rangle,
\eeq
where $k$ is required to be some nonzero integer to ensure the single valuedness 
of the state. Taking the expectation value of ${\cal H}_{\rm NR}$ for the trial state 
$|\Omega' \rangle$, one finds that the energy is shifted as
\beq
\label{E_angular}
{\cal E}'_{\rm NR} = {\cal E}_{\rm NR}^{\rm min} + {2\pi k} \langle j_{\rm NR} \rangle + 
\frac{\pi k^2}{m r}\langle n \rangle.
\eeq
Because ${\cal E}'_{\rm NR} \geq {\cal E}_{\rm NR}^{\rm min}$ by definition of 
${\cal E}_{\rm NR}^{\rm min}$, one must have the following inequality for any integer $k$:
\beq
k \langle j_{\rm NR} \rangle + 
\frac{k^2}{2 m r}\langle n \rangle \geq 0.
\eeq
The necessary and sufficient condition for this is
\beq
\label{j_bound}
|\langle j_{\rm NR} \rangle| \leq \frac{\langle n \rangle}{2mr}\,.
\eeq
Integrating over the area of the ring, $S = 2 \pi r \Delta r$, we get
\beq
\label{J_bound}
\frac{|\langle J_{\rm NR} \rangle|}{N} \leq \frac{1}{2mr}\,.
\eeq
So $\langle J_{\rm NR} \rangle/N \rightarrow 0$ in the thermodynamic limit
($r \rightarrow \infty$ with $\Delta r$ fixed), and the circulating current is 
thermodynamically negligible in the ground state. This is the no-go theorem 
for circulating currents \cite{Bohm, Ohashi}.

We note that persistent currents in a {\it mesoscopic} normal-metal ring, driven by 
an external magnetic flux $\Phi$ \cite{Buttiker} do {\it not} constitute a contradiction 
with this theorem. Indeed, the magnitude of total electric current in the ring with 
circumference $L=2\pi r$ is estimated as \cite{Imry}
\beq
\label{eq:mesoscopic}
|\langle J_{\rm mes} \rangle| = -L\frac{dE_{\rm NR}}{d\Phi}
\sim v_{\rm F},
\eeq
where $v_{\rm F}\sim N/(mL)$ is the Fermi velocity; this is the same order as the 
upper bound in Eq.~(\ref{J_bound}) and $|\langle J_{\rm mes} \rangle|/N \rightarrow 0$ 
in the thermodynamic limit $L \rightarrow \infty$. 

In contrast, ``persistent currents'' in a {\it macroscopic} superconducting ring 
are not actually in the ground state, but in the metastable state \cite{Bohm, Ohashi}; 
it can in principle decay into the genuine ground state with no circulating current 
(which has a lower energy), but its lifetime is so long that it can be regarded as 
quasiequilibrium.

\section{Gauge symmetry and extension of Bloch theorem}
\label{sec:extension}
One can ask how general the Bloch theorem is and if it is also applicable to 
relativistic systems, boson systems, systems in electromagnetic fields, 
and so on. In the above proofs, what we made use of is not actually the 
details of the Hamiltonian, but is just the {\it gauge symmetry}. 
Guided by the consequence of the gauge symmetry, one can extend it to 
\emph{generic} systems.

To see it more clearly, we consider a general Hamiltonian density of (charged 
or neutral) fermions, ${\cal H}(\psi)$. We denote the corresponding Lagrangian 
density as ${\cal L}(\psi)$. Our argument can easily be generalized to 
multicomponent fermions, $\psi_i$ ($i=1,2,\dots,N$), and to charged scalar 
fields, $\phi$. For the sake of simplicity, we shall consider the single-component 
fermion, $\psi$.

\subsection{Generalized no-go theorem for total currents}
\label{sec:extension_bloch}
Let us first prove the generalized Bloch-type no-go theorem for total currents.
We assume the existence of the ground state $|\Omega \rangle$ which has the 
lowest ground-state energy, $\langle H \rangle = E_{\rm min}$, and carries a 
nonvanishing total current, $\langle {\bm J} \rangle \neq 0$. Here the total 
particle number current is defined by
\beq
\label{J}
{\bm J} = \int d^3 {\bm x}\, {\bm j}(\bm x),
\eeq
where 
\beq
\label{Noether}
{\bm j}= \frac{\d \cal L}{\d ({\bm \nabla}\psi)}\frac{\delta \psi}{\delta \theta}+{\rm H.c.}
\eeq
is the Noether current associated with the global 
$\U(1)$ particle number symmetry, $\psi \rightarrow e^{i\theta}\psi$. The Noether 
theorem ensures that ${\bm \nabla}\cdot{\bm j}=0$ in the static limit.

Let us consider the total energy for the trial state $|\Omega' \rangle$ defined by 
Eq.~(\ref{GS}), $\langle\Omega'|H(\psi)|\Omega' \rangle$. This is equivalent to the 
total energy for the Hamiltonian in terms of the {\it new field},
\beq
\label{momentum}
\psi'({\bm x}) = e^{i \delta {\bm p}\cdot {\bm x}} \psi({\bm x}),
\eeq
in the {\it ground state}, $\langle\Omega|H(\psi')|\Omega \rangle$.
Here we assumed that the kinetic term is bilinear in $\psi$ and the interaction 
term is invariant under Eq.~(\ref{momentum}). 

The point is that Eq.~(\ref{momentum}) is regarded as the ``gauge transformation,''
\beq
\label{gauge}
\psi'(\bm x) =  e^{i \theta(\bm x)} \psi(\bm x),
\eeq
with $\theta(\bm x) = \delta {\bm p} \cdot {\bm x}$. 
By promoting $\theta({\bm x})$ to a general scalar function of ${\bm x}$, one can 
generally show, by following the standard procedure (see, e.g., Ref.~\cite{Schwartz}), 
that the corresponding variation of the Hamiltonian density is given by
\beq
\label{dH}
\delta {\cal H} ={\bm \nabla}\cdot(\theta {\bm j})
={\bm \nabla}\theta \cdot {\bm j}, 
\eeq
to first order in ${\bm \nabla}\theta$. Here ${\bm j}$ is the Noether current in 
Eq.~(\ref{Noether}). We stress that Eq.~(\ref{dH}) takes the unique form
dictated by the symmetry (although the expression of ${\bm j}$ itself depends 
on the details of the Hamiltonian). 

Setting $\theta({\bm x}) = \delta{\bm p} \cdot {\bm x}$, performing the integral 
over space, and taking the expectation value with respect to $|\Omega \rangle$, 
one finds the shift of the total energy as
\beq
\label{E_general}
\delta E = \delta{\bm p} \cdot \langle {\bm J} \rangle + O(\delta {\bm p}^2).
\eeq
This reproduces Eq.~(\ref{E}) to first order in $\delta {\bm p}$ for the nonrelativistic 
Hamiltonian. The form of the first term on the right-hand side of Eq.~(\ref{E_general}) 
is determined solely by the symmetry, while that of the second term may depend 
on the details of the Hamiltonian. As it is sufficient to consider an infinitesimally 
small $|\delta {\bm p}|$ for our purpose, the second term at order $O(\delta {\bm p}^2)$ 
is irrelevant. 
If $\langle \bm J \rangle \neq 0$ in the ground state, the total energy is lowered 
by choosing $\delta {\bm p}$ in the opposite direction as $\langle \bm J \rangle$, 
which then contradicts the original assumption. Therefore, it follows that 
$\langle {\bm J} \rangle = 0$ in the ground state of any system. 

In essence, the (gauged) $\U(1)$ particle number symmetry of a system prohibits 
spontaneous particle number currents in the ground state, independently of the 
form of the Hamiltonian. Note that, in the presence of external static electromagnetic 
fields, we need to consider the Hamiltonian that also depends on the gauge field,
$H(\psi,A_{\mu})$. Because 
$\langle\Omega'|H(\psi,A_\mu)|\Omega' \rangle=\langle\Omega|H(\psi',A_\mu)|\Omega \rangle$
with the gauge field being {\it not} transformed, our argument is directly applicable
to this case as well.

\subsection{Generalized no-go theorem for circulating currents}
\label{sec:extension_bohm}
This Bloch-type no-go theorem can also be generalized to circulating currents 
in general systems. We consider a ring with the width $\Delta r$ at radius $r$ 
($\Delta r \ll r$) as in Sec.~\ref{sec:bohm}, and consider the total energy for the 
trial state $|\Omega' \rangle$ defined by Eq.~(\ref{GS2}). 
This energy is equal to the one in terms of the new field,
\beq
\label{angular}
\psi'({\bm x}) = e^{i k \phi} \psi({\bm x}),
\eeq
in the ground state, $\langle\Omega|H(\psi')|\Omega \rangle$.
We then regard Eq.~(\ref{angular}) as the gauge transformation (\ref{gauge}) with 
$\theta = k \phi$. We can concentrate on the kinetic term in the $\phi$ direction, 
since the other kinetic and interaction terms in the Hamiltonian remain unchanged 
under this transformation. For general scalar function $\theta(\phi)$, one can show 
that [see Eq.~(\ref{dH})]
\beq
\delta {\cal H}=\frac{1}{r}\frac{\d \theta}{\d \phi}j+O(r^{-2}).
\eeq

Taking $\theta = k \phi$ and performing the line integral in the ground state, 
one finds that the new field in Eq.~(\ref{angular}) shifts the energy ${\cal E}$ as
\beq
\label{E_angular_general}
\delta {\cal E} = 2\pi k \langle j \rangle + O\left(r^{-1} \right).
\eeq
The first term on the right-hand side above reproduces the term in Eq.~(\ref{E_angular}) 
for the nonrelativistic Hamiltonian; again, the form of this term is determined only by 
the gauge symmetry and is universal, regardless of the details of the Hamiltonian. 
In the thermodynamic limit ($r \rightarrow \infty$ with $\Delta r$ fixed), the second term 
at order $O(r^{-1})$ in Eq.~(\ref{E_angular_general}) is irrelevant. To satisfy 
$\delta {\cal E}\geq 0$ for any integer $k$, we must have $\langle j \rangle = 0$. 
This completes the proof of the generalized Bloch theorem.

\section{Applications of generalized Bloch theorem}
\label{sec:application}

\subsection{Application to chiral transport phenomena}
As mentioned in the introduction, for the system of Dirac fermions with 
chirality imbalance in a magnetic field or in a rotation, ``ground-state (or equilibrium) 
currents'' might appear. These chiral magnetic effect (CME) 
\cite{Vilenkin:1980fu, Alekseev:1998ds, Fukushima:2008xe} 
and chiral vortical effect (CVE) \cite{Vilenkin:1979ui, Landsteiner:2011cp} 
are originally computed using field theory in equilibrium as
\beq
\label{CME}
\langle {\bm j}_{\rm CME} \rangle = \frac{1}{2 \pi^2} \mu_5 {\bm B}, \quad
\langle {\bm j}_{\rm CVE} \rangle = \frac{1}{\pi^2} \mu \mu_5 {\bm \omega},
\eeq
respectively. Here ${\bm j}$ is the current density, 
$\mu = (\mu_{\rm R} + \mu_{\rm L})/2$ and $\mu_5 = (\mu_{\rm R} - \mu_{\rm L})/2$ 
are the vector and chiral chemical potentials, ${\bm B}$ is the magnetic field, 
${\bm \omega}$ is the vorticity, and the expectation value is taken in the ground 
state or in equilibrium. It should be remarked that $\langle {\bm j}_{\rm CME} \rangle = 0$ 
is also obtained in some more recent calculations of holography \cite{Rebhan:2009vc,Brits:2010pw,Gynther:2010ed,Jimenez-Alba:2014iia}
and (lattice) field theory \cite{Hou:2011ze,Landsteiner:2012kd,Buividovich:2013hza}.

Assuming Eq.~(\ref{CME}) in the homogeneous system, the total chiral magnetic 
current would be nonvanishing in the ground state. 
However, the generalized Bloch theorem above suggests that such a state is not 
the true ground state; taking the trial state as in Eq.~(\ref{GS}) or (\ref{GS2}) with 
the appropriate $\delta {\bm p}$ or $k$, one could reduce the ground state energy. 
In other words, current carrying ground states do not respect the gauge symmetry 
in Eq.~(\ref{gauge}) (see also Appendix \ref{sec:bc}).

In the context of Weyl semimetals, the absence of the CME in the ground state is 
numerically confirmed in a specific lattice model without interactions \cite{Vazifeh}, 
where the reason is ascribed to the {\it absence of the Lorentz symmetry} in real 
condensed matter systems. However, our argument shows that the nonexistence 
of the ground-state current is, rather, a consequence of the {\it presence of the gauge symmetry}, 
and is general, not limited to the specific Hamiltonian like Ref.~\cite{Vazifeh}; in particular, 
the same result holds even in the systems with the Lorentz symmetry and/or with
interactions. 

Note that one cannot conclude the absence of the CVE in the ground state in 
the thermodynamic limit, unlike the CME. This is because, as noted in 
Ref.~\cite{Vilenkin:1979ui}, the size of a relativistic system in a global rotation 
$\Omega$ cannot be larger than $r=1/\Omega$, above which the velocity of the 
boundary exceeds the speed of light. Hence, the CVE makes sense only in a finite 
system; in this case, the ground-state current can be nonvanishing due to the 
$O(r^{-1})$ term in Eq.~(\ref{E_angular_general}) (see also Sec.~\ref{sec:bohm}).

\subsection{Application to quantum time crystals}
The generalized Bloch theorem for circulating currents can be directly applied to the 
question of (a class of) quantum time crystals (QTC) recently proposed by Wilczek 
\cite{Wilczek:2012jt} (see also Refs.~\cite{STC,Yoshii:2014fwa} for attempts of realization).
The QTC is a hypothetical state of matter that spontaneously breaks the continuous 
translational symmetry in time, analogously to the usual crystals that spontaneously 
breaks the continuous translational symmetry in space. 

As a concrete realization of the QTC, a system that allows for time-dependent 
persistent circulating currents in the ground state of a ring is proposed \cite{Wilczek:2012jt}. 
Recall here that one needs to take the thermodynamic limit to have any spontaneous 
symmetry breaking. However, according to the generalized Bloch theorem above, such 
a current-carrying ground state is prohibited in the thermodynamic limit (although it is 
possible in a finite volume). A similar result was obtained in the language of quantum 
mechanics in Ref.~\cite{Bruno}. This seems also consistent with a more general 
argument for the absence of the QTC \cite{Watanabe:2014hea}.

We remark that the Bloch theorem itself does not exclude a QTC 
characterized by something different from persistent circulating currents.

\section{Physical interpretation of chiral transport phenomena}
\label{sec:physics}

\subsection{Chiral magnetic effect as a nonequilibrium steady current}
\label{sec:Laughlin}
We now explicitly show that the circulating chiral magnetic current can be understood 
as a generalization of the Bloch theorem to a nonequilibrium steady state. Our argument 
is similar to the one by Thouless \cite{Thouless}, which reformulates Laughlin's 
argument for the integer quantum Hall effect (IQHE) \cite{Laughlin} as an extension 
of the argument for the Bloch theorem. To make our discussion clear, we restore 
the units $\hbar$, $c$, and $e$ in this subsection.

\begin{figure}[b]
\begin{center}
\includegraphics[width=3.5cm]{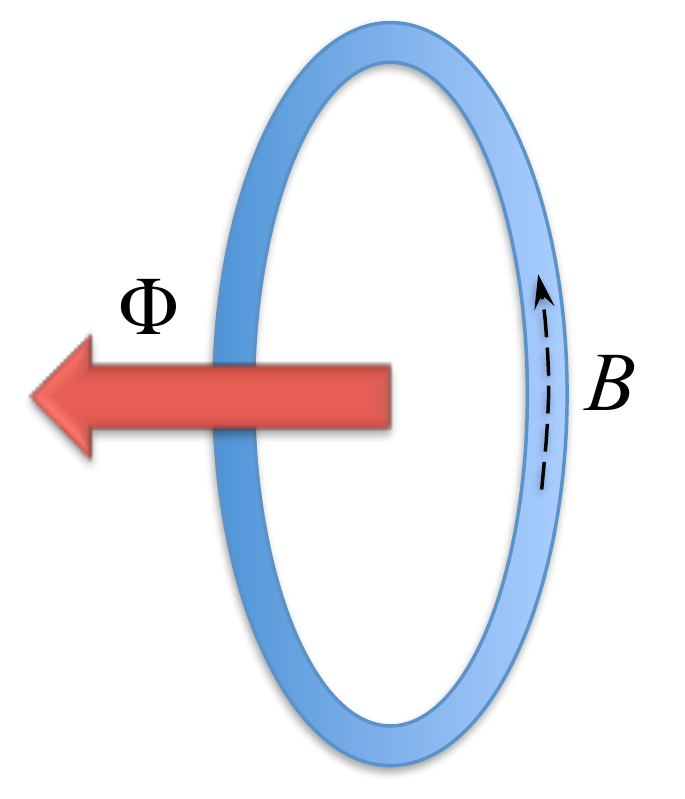}
\end{center}
\vspace{-0.3cm}
\caption{Geometry of the torus pierced by the magnetic field.}
\label{fig:torus}
\end{figure}

We consider noninteracting massless Dirac fermions (right- and left-handed 
massless chiral fermions) in a torus with the cross section $S$ whose inside is 
pierced by a homogeneous magnetic field $B$. This is illustrated in Fig.~\ref{fig:torus}. 
We assume to maintain different chemical potentials $\mu_{\rm R}$ for right-handed 
fermions, and $\mu_{\rm L}$ for left-handed fermions in a torus. We also introduce the 
magnetic flux $\Phi$ threading the hole of the torus.

Let us vary the magnetic flux threading the torus adiabatically 
by one quantum unit,
\beq
\label{gauge_unit}
\delta \Phi \equiv \oint \delta{\bm A} \cdot d{\bm l} =  \frac{2\pi \hbar}{e}\,,
\eeq
where $\delta {\bm A}$ is the change of the gauge field inside the torus. This 
leads to the trivial Aharonov-Bohm phase for the fermions, 
$\exp \left(-i e \delta\Phi/\hbar \right) = 1$, and the system does not change 
from the original state. The only change that can happen is the transfer of 
$N_{\rm B}$ massless fermions from the Fermi surface of left-handed fermions 
to that of right-handed fermions (meaning that the system is in the nonequilibrium 
steady state). In the presence of the chemical potential difference between two 
Fermi surfaces, this transfer requires the energy $N_{\rm B}(\mu_{\rm R}-\mu_{\rm L})$. 
Hence, by the gauge transformation in Eq.~(\ref{gauge_unit}) together with the 
shift of $N_{\rm B}$ fermions, the change of the total energy is given by
\begin{align}
\label{Laughlin}
\delta {E} &= \int d^3{\bm x} \, {\bm j} \cdot \delta{\bm A} - N_{\rm B} (\mu_{\rm R}-\mu_{\rm L})
\nonumber \\
&= I \left(\frac{2\pi \hbar}{e}\right) - N_{\rm B} (\mu_{\rm R}-\mu_{\rm L}),
\end{align}
where $I$ is the current flowing around the torus. As the system comes back to 
the original state, the total energy shift is zero in this process, $\delta {E} =0$. 
We thus get
\beq
\label{jS}
I = \frac{N_{\rm B}e}{2 \pi \hbar}(\mu_{\rm R}-\mu_{\rm L}).
\eeq
This can be viewed as a bulk version of Landauer-type formula \cite{Imry} 
with perfect transmission.

We now determine $N_{\rm B}$. The magnetic field inside the torus gives 
rise to the quantization of energy levels (Landau levels) for Dirac fermions.
The fermions in the lowest Landau level are massless, and the degeneracy per 
unit transverse area is $g_n=eB/(2\pi \hbar c)$; 
the number of gapless modes in the area $S$ is given by $N_{\rm B}=g_n S$.

Substituting it into Eq.~(\ref{jS}), we obtain
\beq
\label{CME_steady}
j = \frac{e^2 \mu_5}{2\pi^2 \hbar^2 c} B,
\eeq
where $j=I/S$ is the current density and $\mu_5 = (\mu_{\rm R} - \mu_{\rm L})/2$.
This is exactly the expression of the CME in Eq.~(\ref{CME}) in the units $\hbar=c=e=1$.
In this way, the CME can be seen as the current of $N_{\rm B}$ {\it bulk} states.

This argument clarifies not only the similarity between the CME and IQHE
via Eq.~(\ref{jS}),%
\footnote{In the original Laughlin's argument for the IQHE \cite{Laughlin}, the 
number of {\it edge} modes moved by the gauge transformation on a ribbon is 
some integer $N_{\rm E}$ (related to the Chern number), and 
$\mu_{\rm R}-\mu_{\rm L}$ is replaced by the voltage between the two edges 
multiplied by the electric charge, $eV$. Then Eq.~(\ref{jS}) reduces to the 
familiar expression of the IQHE, $I={N_{\rm E}e^2}V/h$.} but also the difference 
that the current is carried by bulk (edge) modes for the CME (IQHE).
Equation (\ref{CME_steady}) evades the Bloch theorem, because the system 
is in the nonequilibrium steady state with keeping $\delta E = 0$, similarly to
the IQHE; the current is driven by the ``voltage'' $\mu_{\rm R}-\mu_{\rm L}$.

\subsection{Chiral separation effect as Pauli paramagnetism}
\label{sec:CSE}
So far we have concentrated on the CME at finite $\mu_5$. In the presence of 
$\mu$, the spontaneous {\it axial} current, 
\beq
\label{CSE}
\langle {\bm j}_5 \rangle = \frac{1}{2\pi^2} \mu {\bm B},
\eeq
is also considered to appear. This is called the chiral separation effect (CSE) 
\cite{Son:2004tq, Metlitski:2005pr}. 
Contrary to the vector currents, such as the electric current, there is no axial 
gauge symmetry corresponding to Eq.~(\ref{momentum}) or Eq.~(\ref{angular}). 
This means that the Bloch-type no-go theorem is not directly applicable to the 
CSE and that the total axial current can appear even in the ground state. 
In the following, we shall explicitly show that the CSE is purely a ground-state 
property of relativistic matter---Pauli paramagnetism. For simplicity and convenience, 
we consider a noninteracting relativistic Fermi gas at finite $\mu$. 

The starting point is the Dirac Hamiltonian density,
\beq
\label{Dirac}
{\cal H}_{\rm Dirac} = \psi^{\dag} (-i {\bm \alpha} \cdot {\bm D} - \mu) \psi,
\eeq
where ${\bm \alpha} = \gamma^0 {\bm \gamma}$,
${\bm D} = {\bm \nabla} + i {\bm A}$, and $\psi$ is the four-component Dirac field.
Using the equation of motion, one can rewrite the interaction term
between the gauge field and the current into the form of the ``Pauli term,''
\beq
\label{H_int}
{\cal H}_{\rm int} =  \frac{i}{2\mu} {\bm A} \cdot 
(\psi^{\dag} \overleftrightarrow{\bm \nabla} \psi) 
- \frac{1}{2\mu} \psi^{\dag} {\bm B} \cdot {\bm \sigma} \psi,
\eeq
up to total derivatives. This shows that free \emph{massless} Dirac fermions 
at finite $\mu$ has the magnetic moment $\gamma = e/(2\mu)$ at the tree level
(see also Refs.~\cite{Son:2012zy, Chen:2014cla, Manuel:2014dza}).
This is similar to the magnetic moment for \emph{massive} 
Dirac fermions at $\mu=0$.

Below we take the magnetic field in the $z$ direction, ${\bm B}=(0,0,B)$.
The ``Zeeman splitting" in the second term 
changes the particle energy depending on the spins,
\beq
\label{epsilon}
\delta \epsilon_{{\bm p} \sigma} = -\gamma \sigma_z B.
\eeq
This in turn leads to the change of the distribution functions of fermions,
\beq
\delta n_{{\bm p} \sigma} = \frac{\partial n_{{\bm p} \sigma}}
{\partial \epsilon_{{\bm p} \sigma}}(\delta \epsilon_{{\bm p} \sigma} - \delta \mu),
\eeq
where $n_{{\bm p} \sigma} = \theta (\mu-|{\bm p}|)$.
Because $\delta\mu$ is a scalar quantity, $\delta \mu$ must be an 
even function of $B$, and $\delta \mu \propto B^2$ at the leading order. 
At first order in $B$ (for sufficiently small $B$), 
the variation of the chemical potential $\delta \mu$ is thus negligible.
Then the total number of particle with spin $\sigma$ is given by
\beq
\label{delta_n}
\delta n_{\sigma} = \int \frac{d^3 {\bm p}}{(2 \pi )^3} \delta n_{{\bm p} \sigma}
= \frac{1}{2} N(\mu) \gamma \sigma_z B,
\eeq
where $N(\mu)=\mu^2/\pi^2$ is the density of states at the 
Fermi surface  including spin degrees of freedom.

The axial current is expressed as the net spin polarization,
\beq
\label{j_5}
\langle j^z_5 \rangle = \langle \psi^{\dag} \Sigma^z \psi \rangle 
= \delta n_{\uparrow} - \delta n_{\downarrow}
\eeq
where $\Sigma^ i =  \gamma_5 \gamma^0 \gamma^i$ is the spin operator.
From Eq.~(\ref{delta_n}), this current can be computed as
\beq
\langle j^z_5 \rangle = N(\mu) \gamma B = \frac{1}{2\pi^2} \mu B,
\eeq
which is nothing but the CSE in Eq.~(\ref{CSE}). 

Although the CME and CSE look superficially similar in expressions
(\ref{CME}) and (\ref{CSE}), they are different in that the Bloch theorem is 
applicable to the former, but not to the latter. This is intimately related to the 
presence (absence) of the \U(1) vector (axial) gauge symmetry.

\section{Conclusion}
\label{sec:conclusion}
In this paper, we have extended the Bloch theorem to generic systems based
on the consequence of the gauged \U(1) particle number symmetry. The 
generalized Bloch theorem excludes the possibility of the chiral magnetic effect 
and quantum time crystals as persistent currents in the thermodynamic limit. 
We have also shown that the chiral magnetic effect can be understood as the 
nonequilibrium steady current, similarly to the integer quantum effect.

The crux of the proof of the generalized Bloch theorem for vector currents is the 
\U(1) (vector) gauge symmetry: a coordinate-dependent space translation 
(or spatial rotation) of a state can be regarded as the vector gauge transformation 
of fields in a theory. However, as there is no such \U(1) axial gauge symmetry, 
the Bloch-type no-go theorem is not applicable to the axial current. We have 
explicitly demonstrated that the chiral separation effect is the spontaneous axial 
current in the ground state. It would be interesting to apply our arguments to other 
currents, such as heat currents and spin currents. 

Finally, it should be possible to extend the Bloch-type no-go theorem considered 
in this paper to generic systems at {\it finite} temperature, in a way similar to 
Ref.~\cite{Ohashi}.

\acknowledgments
We thank Yoji Ohashi, Keiji Saito, Atsuo Shitade, and Ryo Yokokura for useful 
discussions. We especially thank Yoji Ohashi for drawing their attention to
Refs.~\cite{Bohm, Ohashi} and critical reading of the manuscript. 
This work was supported by JSPS KAKENHI Grant No. 26887032.

\appendix

\section{Chiral magnetic effect, gauge invariance, and boundary conditions}
\label{sec:bc}

We here provide an alternative explanation (for a specific boundary condition)
based on the gauge invariance that the total chiral magnetic current should 
vanish in the ground state. 
(See also Ref.~\cite{Kharzeev:2013ffa} for a related discussion.)

Substituting the CME in Eq.~(\ref{CME}) into the interaction term between 
the gauge field and the current,
\beq
H_{\rm int}=\int d^3{\bm x} \, {\bm A}\cdot{\bm j},
\eeq
we have
\beq
H_{\rm CS}=\frac{\mu_5}{2\pi^2}\int d^3{\bm x} \, {\bm A}\cdot{\bm B}.
\eeq
This is the effective Chern-Simons term induced at finite $\mu_5$ \cite{Redlich:1984md}.
Note that this is gauge invariant up to surface terms. 
By the gauge transformation, ${\bm A}\rightarrow{\bm A}-{\bm \nabla}\Lambda$ 
with $\Lambda({\bm x})$ being any scalar function, this energy is shifted as
\beq
\Delta H_{\rm CS}=\frac{\mu_5}{2\pi^2}\int_S \Lambda({\bm x}){\bm B}\cdot d{\bm S},
\eeq
where $S$ is the boundary of the region under consideration. To maintain the 
gauge invariance (i.e., $\Delta H_{\rm CS}=0$) for any $\Lambda$, one can 
take the following boundary condition at $S$: 
(i) $\langle{\bm j}\rangle \cdot d{\bm S}=0$, or
(ii) the periodic boundary condition for $\langle{\bm j}\rangle$.

In fact, this requirement is related to the conservation of the particle number, 
and is not limited to the CME. We consider $N$ fermions in a finite (but sufficiently 
large) volume region $V$ with the boundary $S=\d V$. We assume the local 
current conservation, $\d_{\mu}j^{\mu}=0$ with $j^{\mu}$ being the particle 
number current. However, the {\it local} current conservation does not necessarily 
mean the {\it global} charge conservation. Indeed, using the local current conservation, 
one has
\beq
\label{N}
\d_t N = -\int_S \, \langle{\bm j}\rangle \cdot d{\bm S},
\eeq
which can be nonzero unless one chooses the boundary condition at $S$ 
appropriately. In order for $N$ to be conserved in the region $V$, one needs 
to choose the boundary condition (i) or (ii) above.

For the boundary condition (i), one can show that%
\footnote{We here assume that $|\langle{\bm j}\rangle|$ decreases 
faster than $|{\bm x}|^{-1}$ at a sufficiently large distance $|{\bm x}|$.}
\beq
\langle J^i \rangle = \int d^3{\bm x}\, \d_k(x^i j^k)
=\int_S \, x^i \langle j^k \rangle dS^k = 0,
\eeq
which is the same conclusion as the generalized Bloch theorem.
In other words, if $\langle {\bm J} \rangle \neq 0$ in the region $V$, it means that 
$\d_t N \neq 0$, and then the system under consideration would not be static. 
Note that this argument is not limited to the CME or CVE and is applicable 
to any system with the boundary condition (i). This argument, however, cannot 
simply be carried over to the case of the boundary condition (ii) and to circulating 
currents; in those cases, one needs to resort to the Bloch-type argument to show 
vanishing total vector currents in the ground state, as we have shown above.

\end{document}